\documentclass[runningheads]{llncs}
\usepackage[T1]{fontenc}
\usepackage{sidecap}
\usepackage{graphicx}
\usepackage{amsfonts}
\usepackage{amsmath}
\usepackage{amssymb}
\usepackage{array}
\usepackage{booktabs}
\usepackage{longtable}
\usepackage{algorithm}
\usepackage{algorithmic}

\usepackage{todonotes}
\newcommand{\fnm}{\footnotemark}
\newcolumntype{L}[1]{>{\raggedright\let\newline\\\arraybackslash\hspace{0pt}}m{#1\textwidth}}
\newcolumntype{C}[1]{>{\centering\let\newline\\\arraybackslash\hspace{0pt}}m{#1\textwidth}}
\newcolumntype{R}[1]{>{\raggedleft\let\newline\\\arraybackslash\hspace{0pt}}m{#1\textwidth}}

\makeatletter
\@mparswitchfalse 
\makeatother
\normalmarginpar   

\begin{document}
\title{Implementation of QR factorization of tall and very skinny matrices on current GPUs}
\titlerunning{Tall and very skinny QR on current GPUs}
%
%
\author{Jonas Thies\inst{1}\orcidID{0000-0001-9231-9999} \and
Melven R\"ohrig-Z\"ollner\inst{2}\orcidID{0000-0001-9851-5886} 
}
\authorrunning{J. Thies and M. R\"ohrig-Z\"ollner}
%
\institute{Delft University of Technology\\
Delft Institute of Applied Mathematics\\
Mekelweg 4, 2628 CD Delft, The Netherlands\\
\email{J.Thies@TUDelft.nl}
\and
German Aerospace Center\\
Institute of Software Technology\\
Linder H\"ohe, 51147 Cologne, Germany\\
\email{Melven.Roehrig.Zoellner@DLR.de}
}
\maketitle              
\begin{abstract}
We consider the problem of computing a QR (or QZ) decomposition of a real, dense, tall and very skinny matrix. That is,
the number of columns is tiny compared to the number of rows, rendering most computations completely or partially
memory-bandwidth limited. The paper focuses on recent NVIDIA GPGPUs still supporting 64-bit floating-point arithmetic,
but the findings carry over to AMD GPUs as well. We discuss two basic algorithms: Methods based on the normal equations
(Gram matrix), in particular Cholesky-QR2 and SVQB, and the "tall-skinny QR" (TSQR), based on Householder
transformations in a tree-reduction scheme. We propose two primary optimization techniques: Avoiding the write-back of
the Q factor (``Q-less QR''), and exploiting fast local memory (shared memory on GPUs). We compare a straight-forward
implementation of Gramian-based methods, and a more sophisticated TSQR implementation, in terms of performance achieved,
time-to-solution, and implementation complexity. By performance modelling and numerical experiments with our own code and a vendor-optimized
library routine, we demonstrate the crucial need for specialized methods and implementations in this memory-bound to
transitional (memory/compute-bound) regime, and that TSQR is competitive in terms of time-to-solution, but at the cost
of an investment in low-level code optimization.

\end{abstract}

\section{Introduction}
\label{sec:introduction}
We consider the problem of computing a QR-decomposition of a real, dense matrix $X\in \mathbb{R}^{m\times n}$,
where $m \gg n$. Such a matrix is often referred to as ``tall and skinny''. There are many use cases for tall-and-skinny QR, e.g.,
computing an orthogonal basis of a subspace in iterative linear or eigenvalue solvers, as a building block for tensor factorization,
for data fitting or compression.

Formally, we want to compute $Q\in \mathbb{R}^{m \times n}$
and $R \in \mathbb{R}^{n\times n}$ such that $X=QR$, $Q^TQ=I$, and $R$ is upper triangular. 
If $X$ is rank-deficient ($r=\mathrm{rank}(X)<n$), $R$ will have some zero diagonal entries.
Specialized algorithms for the tall and skinny case include methods based on the normal equations (Gram matrix): Variants of 
SVQB~\cite{Stathopoulos2002} and Cholesky-QR\cite{choleskyqr2},
and the ``tall-and-skinny QR'' (TSQR~\cite{demmelTSQR}), which is based on Householder transformations. 
SVQB produces the decomposition $XB=Q, Q^TQ=I$, which can be transformed into the QZ-factorization: $X=QZ, Z=\Sigma U$. 
Here $\Sigma, Q$, and $U$ are the singular values and right (left) singular vectors of $X$, respectively.

The entire class of algorithms is referred to as ``communication-avoiding'' methods as they aim to reduce the data traffic between memory and computing units, and among the nodes of a cluster.
For matrices with relatively few columns, this reduces the time-to-solution because the required floating point calculations do not amortize data movement.

In this paper, we focus specifically on the case of tall and \textbf{very} skinny matrices, where the number of columns is so small that the memory traffic dominates the computing time.
We focus on algorithmic variants for the `Q-less QR', that is, they do not return $Q$ or its generating coefficients at all.
If $X$ is sufficiently well-conditioned, $Q$ can be re-constructed at a later stage if needed. 
The term was introduced by Fan et al.~\cite{Fan2016} and is useful, e.g., for large tensor decomposition algorithms such as TT-SVD~\cite{TTSVD2022}.
It can also be used to solve least squares problems $\min_x \|Ax-b\|_2$ with a single pass over the data by calculating the Q-less QR decomposition of the extended matrix $\begin{pmatrix}A & b\end{pmatrix}$.
Our implementations focus on achieving high performance on current NVIDIA GPUs, with the primary optimization technique being manual caching in shared memory. The results should carry over to other types of GPUs (e.g., AMD) that support double precision arithmetic and provide fast local memory per multiprocessor.

\section{Related work}
\label{sec:related_work}
Numerical results on accuracy and stability of various methods can be found in classical textbooks~\cite{golubvanloan,Higham2002}.
The operations of standard Householder QR can be combined into level-3 BLAS operations, resulting in the fast LAPACK routine \texttt{dgeqrf}.
But unconditional backward stability is only achieved with suitable pivoting~\cite{Cox1998}.
Unfortunately, pivoting reduces the performance significantly as it usually requires bandwidth-limited level-2 BLAS operations (LAPACK routine \texttt{dgeqp3}).
This performance penalty increases for newer CPU generations and GPUs due to the widening gap between memory bandwidth and floating-point performance (the von Neumann bottleneck).
Therefore, researchers try to overcome the costs of pivoting, e.g., by solving multiple shifted linear systems by unpivoted QR\cite{Chevalier2024}, or by skipping small columns in the QR calculation~\cite{SidLakhdar2023}.
These approaches seem promising for specific classes of well-behaved problems, but might not be stable in general.
In particular, even with pivoting, the maximal error for ``truncating'' small diagonal entries in the triangular factor 
of the QR decomposition grows with $O(2^r)$ for rank $r$, see~\cite{Higham1990,Kawamura2021}.
For computing the SVD, one typically employs a preceding unpivoted Householder-QR orthogonalization step for rectangular and tall-skinny matrices, 
see~\cite{Trefethen1997}.

The important case of tall and skinny matrices, where data transfers determine runtime more than anything else,
motivated the development of so-called communication-avoiding algorithms.
TSQR (tall-skinny QR) and CAQR (communication-avoiding QR, which uses TSQR as a panel factorization for not-so-skinny to square matrices)~\cite{demmelTSQR} allow much faster implementations:
TSQR requires logarithmically less data to be transferred and ``messages'' (number of data transfers) compared to Householder QR.
This provides the basis for high-performant QR implementations for today's cache-based multicore CPUs, highly parallel GPUs~\cite{Anderson2011} and clusters of those~\cite{Langou2010,Constantine2011}.
More recent work on implementing CAQR for square matrices on GPUs is presented in \cite{Leng2025,Schofield2019}.
Other work related to TSQR includes
a variant with approximate Householder reflectors~\cite{Tomas2020},
a parallel pivoting strategy~\cite{Demmel2015}, and
combining TSQR with an outer block modified Gram-Schmidt iteration, which is unconditionally backward stable~\cite{Barlow2019}.

A communication-avoiding alternative to TSQR are methods based on the Gram matrix, $C=X^TX$. Such approaches have the inherent disadvantage that
the condition number of the input matrix $X$ is squared before computing the final factorization. On the other hand, the core computation is a tall-and-skinny GEMM, which is a well-optimized kernel on GPUs~\cite{Ernst2021}. SVQB~\cite{Stathopoulos2002}, in fact a tall-and-skinny SVD, was the first publication in the context of avoiding communication, although the underlying principles were known for decades.
Recently, Cholesky-QR has received more attention than SVQB, probably because for larger $n$, the Cholesky factorization is cheaper.
For better robustness, the method is applied twice using re-orthogonalization~\cite{bjorck}, a method that we refer to as CholQR2~\cite{yamamoto2015choleskyqr2}.
The authors of~\cite{Stathopoulos2002} also consider re-orthogonalization in the context of the ``two-phase problem'' ($QR=(I-YY^T)X$) for orthogonalizing $X$ with respect to some already orthogonal matrix $Y$.
Even if CholQR2 theoretically has twice the cost of TSQR, its implementations can be more efficient, as illustrated in \cite{choleskyqr2} for a multicore CPU.
In~\cite{Yamazaki2015}, the Gramian is accumulated in higher precision to alleviate the restriction on the condition number, with similar performance as the ``standard'' CholQR2 algorithm for sufficiently skinny matrices on a multi-GPU system.
To improve the numerical behavior for ill-conditioned matrices $X$, \cite{Fukaya2020} introduces the shifted CholQR3 variant, which is backward stable for $\kappa(X)<\mathcal{O}(\epsilon^{-0.5})$.
The performance of several algorithms (HHQR, TSQR, shifted CholQR3 and CGS2) on CPU clusters is compared in~\cite{Fukaya2023}, finding a clear advantage of CholQR3.
In~\cite{Fukaya2024}, an iterated CholQR with column pivoting is proposed that avoids the second pass.

Finally, methods from randomized numerical linear algebra have been proposed to mitigate pivoting costs in QR factorizations.
Randomized QR with column-pivoting (QRCR~\cite{Duersch2017,Martinsson2017,Duersch2020}) uses a random sampling of the input matrix to estimate column norms. In \cite{Melnichenko2025},  randomization is used to precondition the Gramian of a CholQR algorithm for tall and not-so-skinny matrices.

\section{Methodology}
\label{sec:methodology}
In this section, we will briefly describe several algorithms for
the task at hand: Cholesky-QR2, SVQB2 and (Householder) TSQR.
Based on the Roofline performance model~\cite{Williams2009}, we establish practical performance bounds for each method.
Underlying the Roofline model are the following assumptions: (i) The execution speed of a numerical algorithm is limited by
the slowest path between the data and the execution units and the floating point units performing the required arithmetic operations;
(ii) Data transfers and calculations can be perfectly overlapped.
If the arithmetic intensity of the algorithm (in flops/byte) 
is given by $I$, the bandwidth of the memory in which the data resides is $b$, and the theoretical peak performance of the floating point units is $R_\text{peak}$, then the \textit{Roofline performance} limit is given by
\begin{equation}
R_\text{roof} = \min(R_\text{peak}, I\cdot b).
\end{equation}
We can determine the primary bottleneck for a specific algorithm/hardware combination by defining 
the machine balance $M = \frac{R_\text{peak}}{b}$. If $I<M$, the algorithm is called \textit{memory-bound},
otherwise it is called \textit{compute-bound}. The model is typically unrealistic near the ``kink'' $M=I$. We therefore
use the mixbench~\cite{Konstantinidis2017} microbenchmark to determine $R_\text{roof}$. Limitations of our model are that
using tensor cores for GEMM-like operations is not taken into account (mixbench uses CUDA cores with fused multiply-adds (FMA)), 
and that the ``perfect overlapping'' assumption is easily violated on GPUs as they lack the out-of-order execution capabilities of modern CPUs, and latencies
are relatively large, making them harder to hide by computation.

\subsection{CholQR2 and SVQB2}

These algorithms first compute the Gram matrix $C=X^TX$. As this makes them unsuitable for $\kappa_2(X)>\sqrt{\epsilon_{mach}}$),
we will always apply the algorithm twice for increased robustness. Nevertheless, the resulting methods are only suitable for moderately
ill-conditioned matrices. A ``Q-less'' CholQR2 algorithm without additional stabilization is sketched in Alg.~\ref{alg:Qless_CholQR2}.

\begin{algorithm}[h!]
\caption{Q-less Cholesky-QR2\label{alg:Qless_CholQR2}}
\begingroup
\footnotesize
\begin{algorithmic}[1]
\REQUIRE{$X \in \mathbb{R}^{m \times n}, m\gg n$}
\ENSURE{$R: (XR^{-1})^T(XR^{-1})=I$}
\STATE{$C \gets X^TX$} \COMMENT{\texttt{tsmttsm}, device) \label{tsmttsm}}
\STATE{$R_1 \gets \mathrm{chol}(C)$} \COMMENT{host\label{step:chol1}}
\STATE{$C \gets (X/R_1)^T(X/R_1)$} \COMMENT{\texttt{tsmRttsmR}, device \label{tsmRttsmR}}
\STATE{$R_2 \gets \mathrm{chol}(C)$ (host)\label{step:chol2}}
\STATE{$R \gets R_2 \cdot R_1$}
\end{algorithmic}
\endgroup
\end{algorithm}

For the performance model, we will ignore steps that do not involve $m$, as we are interested in the case of very few columns, $n\lessapprox 100$).
\begin{itemize}
    \item \texttt{tsmttsm} loads $m\cdot n$ doubles and performs $2mn^2$ flops, resulting in a computational intensity of $I_{tsmttsm}=\frac{n}{4}$ flops/byte.
    \item \texttt{tsmRttsmR} loads $m\cdot n$ doubles and performs $3mn^2$ flops, resulting in an arithmetic intensity of $I_{tsmRttsmR}=\frac{3n}{8}$ flops/byte.
    \item Operations involving only $n\times n$-matrices are neglected. In the experiments,
    the two "tall \& skinny" kernels account for more than 98.7 percent of the total elapsed time
    for the larger two runs (see Table~\ref{tab:runtimes}).
\end{itemize}
A limitation of the model is that the data dependency within a thread block during the triangular solve is ignored.

For the SVQB2 algorithm, the Cholesky factorization in steps~\ref{step:chol1} and~\ref{step:chol2} is replaced by an eigenvalue decomposition,
$C=U\Lambda U^T$, and the triangular solve in step~\ref{tsmRttsmR} is replaced by a matrix-matrix multiplication with $D=U\Lambda^{-1/2}$.
Pivoting is inherent to this approach, and it facilitates handling rank-deficiency or truncating small singular values by replacing corresponding entries by 0 on the diagonal of $\Lambda^{-1/2}$.
The number of flops in step~\ref{tsmRttsmR} increases from $3mn^2$ to $4mn^2$, giving the arithmetic intensity $I_{tsmmttsmm}=\frac{n}{2}$ flops/byte. Other than that, SVQB2 costs the same: According to the Roofline model,
the increased robustness over CholQR2 practically comes for free for the case of tall and \textit{very} skinny matrices. Eliminating the triangular solves
is expected to increase the predictive power of the performance model as well.

Our implementation of these
 algorithms is based on NVIDIA Warp~\cite{Macklin2022}, which provides tile primitives for matrix-matrix products and triangular solves in shared memory.
 A fixed number of threads and a grid-stride loop are used to reduce the number of atomic adds during the reduction.

\subsection{Householder TSQR}

Our implementation consists of two stages:
First, $k$ thread blocks each factor a small block of $m/k$ rows of $X$ in shared memory, and write the resulting $R_i$ back to
global memory to form $Y\in \mathbb{R}^{kn\times n}$. Then, a single thread block factors $Y$
to produce the final triangular factor $R$, as shown in Algorithm~\ref{alg:Qless_HouseholderTSQR_outerCalls}.
\begin{algorithm}[ht!]
\caption{Calling the Q-less Householder TSQR kernels on the GPU\label{alg:Qless_HouseholderTSQR_outerCalls}}
\begingroup
\footnotesize
\begin{algorithmic}[1]
    \REQUIRE{$X \in \mathbb{R}^{m \times n}, m{\gg}n$, number of GPU thread blocks $k$}
    \ENSURE{$R\in\mathbb{R}^{n\times n}: QR = X$ for some orthogonal matrix $Q$}
    \STATE{$\begin{pmatrix}Y_1\\ \vdots\\Y_k\end{pmatrix}\gets \begin{pmatrix}Q_{1,1}^T \\ & \ddots \\ && Q_{1,k}^T\end{pmatrix}\begin{pmatrix}X_1\\ \vdots\\X_k\end{pmatrix}$} \COMMENT{\texttt{Qless\_hhTSQR}$\langle k \rangle$, device}
    \STATE{$R\gets Q_2^T Y$} \COMMENT{\texttt{Qless\_hhTSQR}$\langle 1 \rangle$, device}
\end{algorithmic}
\endgroup
\end{algorithm}

The small hhTSQR factorization in shared memory is presented in Algorithms~\ref{alg:hhTSQR1} and~\ref{alg:hhTSQR2}.
The number of thread blocks $k$ and inner block size $b$ are chosen based on hardware properties, most importantly the
number of SMs and the amount of shared memory available per thread block.

\begin{algorithm}[h!]
\caption{\texttt{Qless\_hhTSQR} GPU kernel (per thread block)\label{alg:hhTSQR1}}
\begingroup
\footnotesize
\begin{algorithmic}[1]
    \REQUIRE{$X \in \mathbb{R}^{(m/k) \times n}, m>>n$, block size $b$}
    \ENSURE{$R\in\mathbb{R}^{n\times n}: QR = X$ for some matrix of orthogonal columns $Q$}
    \STATE{$R \gets 0$} \COMMENT{shared memory}
    \STATE{$W \gets X_{1:b,:}$} \COMMENT{asynchronously copy to shared memory}
    \FOR{$i=1,\dots, m/k/b$}
        \STATE{$\begin{pmatrix}W\\ R\end{pmatrix} \gets Q_i^T \begin{pmatrix}W\\R\end{pmatrix}$} \COMMENT{in-place trapezoidal HHQR, overwrites upper part of $W$}
        \IF{$i+1 < m/k/b$}
            \STATE{$R \gets W_{1:n,:}$} \COMMENT{move data in shared memory}
            \STATE{$W \gets X_{(i+1)b+1:(i+2)b,:}$} \COMMENT{asynchronously copy to shared memory}
        \ENDIF
    \ENDFOR
    \STATE{$R\gets W_{1:n,:}$} \COMMENT{copy to global memory}
\end{algorithmic}
\endgroup
\end{algorithm}
\begin{algorithm}[ht!]
\caption{In-place trapezoidal HHQR kernel\label{alg:hhTSQR2}}
\begingroup
\footnotesize
\begin{algorithmic}[1]
    \REQUIRE{$\begin{pmatrix}W\\R\end{pmatrix} \in \mathbb{R}^{(b+n) \times n}$}
    \ENSURE{$W'_{1:n,1:n} = Q^T\begin{pmatrix}W\\R\end{pmatrix}$ for some matrix of orthogonal columns $Q$}
    \FOR{$i=1,3,5,\dots,n$}
        \STATE{Calculate $W_{i,i}$ and Householder vector $v$ for column $i$}
        \STATE{Apply $v$ to column $i+1$}
        \STATE{Calculate $W_{i+1,i+1}$ and Householder vector $w$ for column $i+1$}
        \STATE{Apply $(v,w)$ to columns $i+2,\dots,n$ in steps of 4.\label{2x4_microkernel}}
    \ENDFOR
\end{algorithmic}
\endgroup
\end{algorithm}

For the performance model, we again consider GPU global memory transfers and double-precision (FMA) floating-point operations (tensor cores are not used),
and drop terms that do not involve $m$.
Algorithm~\ref{alg:Qless_HouseholderTSQR_outerCalls} needs to transfer the complete matrix $X$ once from global memory to shared memory for the 
first (parallel) \texttt{Qless\_hhTSQR}, i.e., load $mn$ doubles. Memory traffic associated with the $kn^2$ elements of $Y$ are ignored.
A lower bound for the number of FP64 floating-point operations is $2mn^2$, where we only count the
$m/b$ trapezoidal HHQR steps with $2(b+1)n^2$ Flops each.
Some operations in the reduction and the second Householder TSQR call (requiring only $O(n^3)$ operations) are ignored.
This yields a computational intensity of $I_\text{c,TSQR} = \frac{n}{4}$ Flop/Byte.
The main limitation of the model is in Alg.~\ref{alg:hhTSQR2}, line~\ref{2x4_microkernel}. This is a tiny product of a $2\times b$ and a $b\times 4$ matrix,
where the former resides in registers and the latter in shared memory. To determine an accurate performance upper bound, shared memory latency and bandwidth
would have to be taken into account, but this is beyond the scope of this study.
In summary, our performance model for TSQR is expected to be most accurate for very large $m$, and too optimistic in particular for (near) compute-bound cases, i.e., larger $n$.

\section{Numerical Experiments}
\label{sec:experimental_results}
We are now ready to put our proposed implementations to the test.
To keep the discussion clear, we focus on a single GPU type (the NVIDIA H100).
As can be seen in Table~\ref{tab:gpu_hardware}, the high-end GPUs currently on the market
are qualitatively similar in the relevant stats for this work, i.e., HBM bandwidth,  shared memory size, and DP flop rates.
For the memory bandwidth, instead of the theoretical, we use $b=2.15 GB/s$, which was the maximum value measured by mixbench.

\begingroup
\begin{table}[h!]
\setlength{\tabcolsep}{2.5pt}
\footnotesize
\begin{tabular}{@{}llllllll@{}}
\toprule
GPU Model    &  SM/CU &   Shared Mem & HBM      &  HBM      & FP64 FMA & FP64 \\
             & count\fnm[1]  &   per SM/CU     & capacity & bandwidth & (vector) & (tensor) \\
\midrule
NVIDIA H100  &        132	&       228 KB &        80 GB          & 3.4 TB/s   & 34 TFlop/s & 67 TFlop/s \\ 
NVIDIA B100  &     160 &	228 KB\fnm[2] & 192 GB         & 8.0 TB/s & 30 TFlop/s  & 40 TFlop/s \\
\midrule
AMD MI300X   & 304 &       64 KB\fnm[3] &     192 GB              & 5.3 TB/s   & 82 TFlop/s & 163 TFlop/s \\ 
AMD MI350X   & 256 &       64 KB\fnm[3] &     288 GB              & 8.0 TB/s & 72 TFlop/s  & 144 TFlop/s \\ 
\bottomrule
\end{tabular}
\caption[Relevant hardware characteristics of recent GPUs from NVIDIA and AMD]
{Relevant hardware characteristics of recent GPUs from NVIDIA\fnm[4] and AMD\fnm[5].\label{tab:gpu_hardware}}
\end{table}
\endgroup

\begingroup
\scriptsize
\raggedright
\footnotetext[1]{NVIDIA B100 is typically power-limited to 700W compared to the 1000W B200.}
\footnotetext[2]{NVIDIA Blackwell adds a separate 256 KB Tensor Memory (TMEM) per SM alongside the 228 KB Shared Memory.
This is not utilized in this paper.}
\footnotetext[3]{AMD uses Compute Units (CUs). In CDNA 3/4, each CU typically has 64 KB of Local Data Share (LDS), which is the AMD equivalent of Shared Memory.}
\footnotetext[4]{NVIDIA Corporation. (2025). NVIDIA data center technologies website.
\url{https://www.nvidia.com/en-us/data-center/h100/}, resp.
\url{https://www.nvidia.com/en-us/data-center/technologies/blackwell-architecture/}.
}
\footnotetext[5]{
Advanced Micro Devices, Inc. (2025). AMD Instinct specifications. 
\url{https://www.amd.com/en/products/accelerators/instinct/mi300/mi300x.html}, resp.
\url{https://www.amd.com/en/products/accelerators/instinct/mi350/mi350x.html}.}
\endgroup

\subsection{Experimental setup}

We construct the matrix $X\in\mathbb{R}^{m\times n}$ such that $n=[1, 8, 16, 32, 64]$, keeping $mn$ constant
(i.e., the memory footprint of $X$ is kept constant while increasing the number of columns and thus the computational intensity).
The condition number $\kappa_2(X)$ is kept smaller than $\epsilon_{DP}^{-0.5}$ so that one pass of unpivoted Householder QR, resp.\ two passes
of a Gramian-based method like (CholQR2 or SVQB2) result in a stable and accurate solution. Before running a benchmark, a few
``warm-up'' runs are performed. This means that $X$ already resides on the device, just-in-time compilation is not included in the measurement,
etc. At least 50 runs are performed for each case, and the average runtime is reported.

\subsection{Householder QR and the vendor library cuSOLVER}

Our first experiments seek to establish that there is actually a need for specialized algorithms and implementations
for the tall and very skinny QR factorization on the target platform. Table~\ref{tab:standard_QR_runtimes} compares the runtime
of a CUDA (v13.1) implementation of unpivoted Householder QR and the cuSOLVER implementation of LAPACK's \texttt{dgetrf}, as shipped in NVHPC v26.1 for three different problem sizes $mn$.
The Roofline model time here assumes that $X$ is read once from HBM, and $Q$ is written back to HBM. All cases are memory-bound in theory.

\begin{table}[h!]
\centering
\begingroup
\setlength{\tabcolsep}{5pt}
\footnotesize
\begin{tabular}{ccccc}
\toprule
$m$ & $n$ & ``standard'' HHQR  & cuSOLVER \texttt{dgetrf} & Roofline [ms]\\
\midrule
$8.192\cdot 10^6$ &  8 & 3.1 & 89.3  & 0.48 \\
$4.096\cdot 10^6$ & 16 & 14.6 & 90.9 & 0.48 \\
$2.048\cdot 10^6$ & 32 & 57.3 & 8.3  & 0.48 \\
$1.024\cdot 10^6$ & 64 & 214.6 & 9.9 & 0.48 \\
\midrule
$8.192\cdot 10^7$ &  8 & 52.6 & 889  & 4.8 \\
$4.096\cdot 10^7$ & 16 & 105.2 & 917 & 4.8 \\
$2.048\cdot 10^7$ & 32 & 256.1 & 75.7 & 4.8 \\
$1.024\cdot 10^7$ & 64 & 1690 & 86.4 &  4.8 \\
\midrule
$8.192\cdot 10^8$ &  8 & 545 & --  & 48 \\
$4.096\cdot 10^8$ & 16 & 1098 & -- & 48 \\
$2.048\cdot 10^8$ & 32 & 2267 & 752 & 48 \\
$1.024\cdot 10^8$ & 64 & 5700 & 853 & 48 \\
\bottomrule
\end{tabular}
\endgroup
\caption{\label{tab:standard_QR_runtimes} Runtime (in milliseconds) for standard Householder QR (not TSQR) without pivoting (own implementation and \texttt{cusolverDnXgeqrf}).
Both routines calculate the orthogonal matrix $Q$ in implicit form. 
The cases marked with ``--'' cannot be run with cuSOLVER due to its 32-bit indexing restriction.}
\end{table}

From Table~\ref{tab:standard_QR_runtimes} we see that a straight-forward implementation can outperform the vendor library for very skinny cases ($n\le16$), but neither implementation gets anywhere close to the runtime predicted by the Roofline model.

\subsection{CholQR2 and SVQB2}

\begin{figure}[h!]
\includegraphics{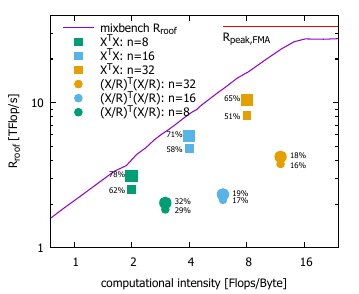}
\hfill
\includegraphics{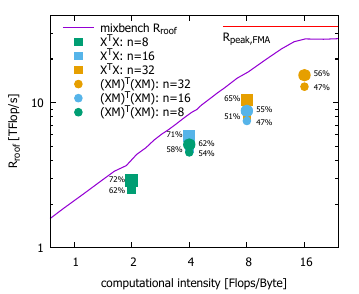}
\caption{\label{fig:cholQR2_h100} Percentage of the Roofline performance $R_\text{roof}$ achieved by the two GPU kernels in CholQR2 (left) and SVQB2 (right),
for varying values of the number of columns $n$. The number of rows $m$ corresponds to the smallest (largest) set of experiments in 
Table~\ref{tab:runtimes}, indicated by small (large) markers, respectively.}
\end{figure}

The performance of the Gramian-based methods is dominated by the \texttt{tsmttsm} kernel (computing $C=X^TX$) and either of the fused kernels
\texttt{tsmRttsmR} ($C=(X/R)^T(X/R)$ or \texttt{tsmmttsmm} ($C=(XB)^T(XB)$ that combine a tall-skinny matrix multiplication with either a small triangular solve (CholQR2)
or a small matmul (SVQB2). As can be seen from Fig.~\ref{fig:cholQR2_h100}, the latter achieves a higher fraction of the Roofline performance as it avoids
the tile-wise triangular solves, which are executed sequentially by a thread block.

\subsection{TSQR}

\begin{SCfigure}[0.6][h!]
\centering
\includegraphics{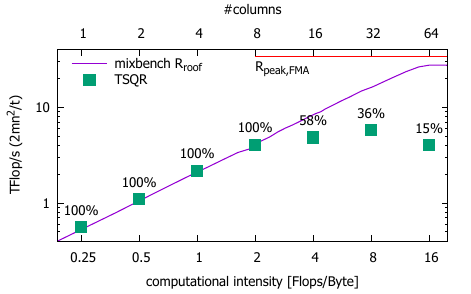}
\caption{\label{fig:tsqr_h100} Percentage of the Roofline performance $R_\text{roof}$ achieved by our TSQR implementation for varying numbers of columns $n$.}
\end{SCfigure}

The Q-less TSQR requires only a single load of the matrix $X$, so in the purely memory-bound regime it can theoretically be twice as fast as CholQR2 or SVQB2.
On the other hand, its kernels are not GEMM-based, so achieving high performance is much more challenging.
Fig.~\ref{fig:tsqr_h100} shows that our implementation achieves ideal Roofline performance for $n\le8$.
For $n > 8$, synchronization effects, instruction latencies and shared memory transfers limit the performance to roughly 6~TFlop/s ($n=32$).
For $n > 32$, the performance decreases due to smaller block sizes (more synchronization overhead) needed to fit into the shared memory.

\subsection{Comparison of the methods}
Table~\ref{tab:runtimes} lists timing results of CholQR2, SVQB2 and TSQR on the H100 to allow a direct comparison.
Standard ``global'' Householder-QR and the implementation by NVIDIA in cuSOLVER (as we saw in Table~\ref{tab:standard_QR_runtimes}) are very slow for the considered tall-and-very-skinny cases, and easily outperformed by more suitable algorithms.
\begin{table}[h!]
\centering
\begingroup
\setlength{\tabcolsep}{5pt}
\footnotesize
\begin{tabular}{ccccccc}
\toprule
     && CholQR2    &              & SVQB2 &   & Q-less TSQR\\
\midrule
 $m$  & $n$ & $t_{mean}$ & $t_{other}$  & $t_{mean}$    & $t_{other}$ & $t_{mean}$ \\
$8.192\cdot 10^6$ &  8 &    1.384 &   0.1111 &    1.051 &   0.1732 & 0.33 \\
$4.096\cdot 10^6$ & 16 &    2.021 &   0.1162 &    1.204 &   0.2097 & 0.62 \\
$2.048\cdot 10^6$ & 32 &    2.306 &   0.1291 &    1.484 &   0.3188 & 1.8 \\
$1.024\cdot 10^6$ & 64 &      -   &      -   &      -   &      -   & 10.8 \\
\midrule
$8.192\cdot 10^7$ &  8 &     11.2 &   0.1448 &    7.924 &   0.2102 & 2.5 \\
$4.096\cdot 10^7$ & 16 &    17.39 &   0.1644 &    8.728 &   0.2503 & 4.5 \\
$2.048\cdot 10^7$ & 32 &    19.27 &   0.1783 &    9.957 &    0.352 & 8.1 \\ 
$1.024\cdot 10^7$ & 64 &      -   &      -   &      -   &      -   & 28.8 \\
\midrule
$8.192\cdot 10^8$ &  8 &    110.9 &   0.3276 &     77.3 &   0.3921 & 25.8 \\
$4.096\cdot 10^8$ & 16 &      171 &   0.3444 &    83.83 &   0.4427 & 43.4 \\
$2.048\cdot 10^8$ & 32 &    189.2 &    0.357 &    95.19 &   0.5764 & 72.5 \\
$1.024\cdot 10^8$ & 64 &      -   &      -   &      -   &      -   & 209.7 \\
\bottomrule
\end{tabular}
\endgroup
\caption{\label{tab:runtimes} Runtime (in milliseconds) for CholQR2 and SVQB2. $t_{other}$ includes the time needed for the small matrix factorizations on the CPU, latencies introduced by Python and launching kernels, etc.}
\end{table}
Interestingly, the SVQB2 variant outperforms CholQR2 by almost a factor of two as the underlying fused kernel \texttt{tsmmttsmm} leverages more parallelism than the triangular solves in \texttt{tsmRttsmR}.
Overall, SVQB2 achieves a high fraction of the Roofline performance ($\gtrsim50$\%) with a relatively straight-forward implementation.
In contrast, our sophisticated TSQR implementation yields 100\% Roofline performance for very few columns ($n\le 8$), which is thus at least two times faster than SVQB2 in this regime.
For more columns ($n=32$), SVQB2 needs about 130\% of the runtime of the TSQR implementation.
With a more sophisticated implementation, SVQB2 could outperform TSQR in this case, as TSQR features less inner parallelism and is thus difficult to implement efficiently.

\subsection{Beyond 32 columns}

\begin{SCfigure}[0.6][h!]
\includegraphics{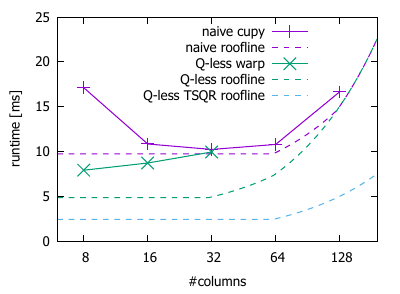}
\caption{\label{fig:warp_vs_cupy}
\thinmuskip=0mu
\medmuskip=0mu
\thickmuskip=0mu
Comparison of the Q-less SVQB2 implementation using warp, and a `naive' implementation using CuPy.
The Roofline limit for the Q-less variant considers $t_{X^TX}+t_{(XM)^T(XM)}$, and the naive variant $t_{X^TX}+t_{XM}+t_{X^TX}$.
Q-less TSQR theoretically needs half of the memory transfers and a third of the Flops.}
\end{SCfigure}

Above we have shown implementations and results for the memory-bound case of up to 32 columns.
The key idea here was to reduce memory traffic by exploiting shared memory on the GPU, and by not explicitly
computing the $Q$ matrix (resp., storing the Householder reflectors). This approach falls short for
larger number of columns, for two reasons: First, the advantage of the ``Q-less'' variant shrinks as the matrix operations
involved become more compute bound. This is illustrated in Fig.~\ref{fig:warp_vs_cupy}, where we compare our Q-less SVQB2
variant to a simple CuPy implementation that stores $Q$ once (as input for the re-orthogonalization step), but not in the second pass.
%
From Fig.~\ref{fig:warp_vs_cupy} it can be concluded that the CuPy implementation achieves very good Roofline performance for $n>=32$.
Furthermore, for larger cases, the underlying CuBLAS operations could use tensor cores on the H100 which is not included here.

The second issue with the Q-less SVQB2 for $n>32$ on current GPUs is that it relies heavily on matrix products in shared memory.
While tiling works well for single GEMM's (or SYRK's, as in our case of $C=X^TX$), it does not carry over to triple products, as required by
our fused kernel $(XB)^T(XB)$.


Our TSQR implementation on the H100 can handle up to 64 columns with the given shared memory capacity.
Theoretically, Q-less TSQR would need fewer operations for $n>64$ than SVQB2 (and CholQR2).
But we do not see a feasible way to implement it given the constraints of current GPUs.
So for larger $n$, non Q-less algorithmic variants like block Gram-Schmidt become useful.

\section{Summary and Conclusions}
\label{sec:conclusions}
We have analyzed the performance of several algorithms for tall-skinny QR factorizations based on the Roofline performance model for an NVIDIA H100 GPU.
Our focus lies on very tall-skinny cases ($n \le 64$ columns), where the problem is inherently memory-bound.
As an optimization, we avoid storing the matrix $Q$ either explicitly or implicitly to reduce data transfers (``Q-less'' QR).
Our results show that this optimization yields a significant speed-up for $n<32$ columns.

More specifically, we tested Q-less variants of the algorithms TSQR, CholQR2 (Cholesky-QR applied twice), and SVQB2 (SVQB applied twice).
Among these, theoretically, TSQR requires the fewest data transfers and operations.
Our sophisticated TSQR implementation yields the fastest results with a speed-up of 3 (for $n=8$ columns) to 1.3 ($n=32$) compared to SVQB2.
SVQB2 is the second-fastest algorithm, and still significantly faster than CholQR2 as the latter requires triangular solves, which are inherently less parallel than pure matrix-matrix multiplications.
Both (SVQB2 and CholQR2) implementations use kernel-fusion to implement, e.g., $(XB)^T(XB)$ in one step, reducing data transfers.
We show that this becomes less relevant for larger numbers of columns $n\ge 64$, where the algorithms become compute-bound, and a straight-forward implementation of SVQB2 provides good performance.

Unfortunately, the TSQR algorithm is limited by the amount of shared memory in the GPU, practically restricting its use to $n\le 32$ columns.
Overall, we achieve a speed-up of 10 for $n=32$ columns (and $\gg 300$ for $n=8$) compared to the vendor library cuSOLVER.
So, we see the need for better implementations of QR algorithms for tall-skinny matrices.
As a trade-off between portability and performance, SVQB2 seems most promising.

\begin{credits}
\subsubsection*{Acknowledgements.}
The authors gratefully acknowledge the scientific support and HPC resources provided by the Erlangen National High Performance Computing Center (NHR@FAU) of the Friedrich-Alexander-Universität Erlangen-Nürnberg (FAU).
The hardware is funded by the German Research Foundation (DFG).

Part of this research was made possible by the Quantum Computing Initiative of the German Aerospace Center (DLR) and the German Federal Ministry for Research, Technology and Space, see \url{qci.dlr.de/projects/QuTeNet}.
\subsubsection*{Disclosure of Interests.}
The authors have no competing interests to declare that are relevant to the content of this article. 
\end{credits}
%
%
%
%
\bibliographystyle{splncs04}
\bibliography{parallel_tsqr_on_cpus_and_gpus}

\end{document}